\documentclass[prl,showpacs,twocolumn]{revtex4}
\usepackage{natbib}
\usepackage{graphicx, epsfig}
\usepackage{color}
\usepackage[debug]{psfrag}
\definecolor{dgreen}{rgb}{0,0.6,0}
\definecolor{gold}{rgb}{1,0.5,0}
\usepackage{psfrag}
\begin{document}

\title{The statistics of particle velocities in dense granular flows}

\author{Sudheshna Moka$^1$ and Prabhu R. Nott}
\affiliation{Department of Chemical Engineering, 
Indian Institute of Science, Bangalore 560 012, India\\
${}^1$Currently at Tata Consultancy Services, Chennai, India}

\begin{abstract}
We present measurements of the particle velocity distribution in the flow of granular material through vertical channels.  Our study is confined to dense, slow flows where the material shears like a fluid only in thin layers adjacent to the walls, while a large core moves without continuous deformation, like a solid.  We find the velocity distribution to be non-Gaussian, anisotropic, and to follow a power law at large velocities.  Remarkably, the distribution is identical in the fluid-like and solid-like regions.  The velocity variance is maximum at the core, defying predictions of hydrodynamic theories.  We show evidence of spatially correlated motion, and propose a mechanism for the generation of fluctuational motion in the absence of shear.

\end{abstract}

\pacs{45.70.Mg, 47.55.Kf \hfill}

\maketitle

	Granular flows are ubiquitous in nature (rock and snow avalanches, movement of sand dunes) and in industrial processes (transport of food grains, ores, pharmaceutical powders), yet they are poorly understood.  In the regime of dense, slow flow there is enduring contact between particles, and shear stresses are generated primarily by dry friction \cite{schofield_wroth68,jackson86}.  In this regime, it is often observed \cite{nedderman_laohakul80,mueth_etal00,losert_etal00} that there is fluid-like deformation only in thin layers, while large regions resist continuous deformation despite the action of a finite shear stress, like a solid.  As in molecular fluids, knowledge of the statistics of particle motion will enhance our understanding of the hydrodynamics of granular materials, and provide insight into the microscopic basis of the fluid-like and solid-like response.  Moreover, the marked difference in the way dense granular materials and conventional fluids respond to applied forces prompts the question of whether there is a fundamental difference in their statistical nature.

	Previous experimental studies on the statistics of granular flows have focused mainly on the rapid flow regime, where grains interact through impulsive inelastic collisions \cite{rouyer_menon00,losert_etal99}.  Studies on the statistics of dense, slow flows have been far fewer, perhaps due to difficulties in probing dense and opaque systems.  Measurement of the distribution of the contact forces \cite{howell_etal99_2} in a monolayer of disks reveal an anisotropic and inhomogeneous network of ``force chains'', which has also been corroborated by particle dynamics simulations \cite{radjai_etal99}.  Measurements of particle displacements by an indirect, spatially averaged technique \cite{menon_durian97} and by direct particle tracking near a transparent wall \cite{choi_etal04} show diffusive motion at long time scales, but the studies differ on their inferences of the microscopic mechanism at play - the former concludes that diffusion arises after repeated impulsive collisions between particles, while the latter attribute it to a mechanism of rearrangement in the network of abiding grain contacts.

	In our investigation, the flow of dry, cohesionless, spherical, almost monodisperse glass beads (Fig.~1b) of mean diameter $d_p\!=\!0.8\,$mm through a vertical channel was studied.  A hopper fed the glass beads into the channel, and the flow rate was controlled by the size of the exit slot at the base of the channel (Fig.~1a). The front and back faces of the channel were smooth, transparent glass plates, and the side walls were aluminium bars, which in some experiments were roughened by sticking 80 grit sandpaper on them.  The channel width $2 W$ and the size of the exit slot could be easily varied.  On illuminating the channel from the front, the reflected light from the beads close to the front glass face were visible as bright spots (Fig.~1b).  Video images of the flowing beads were taken with a CCD camera (mounted on a translating stage) at 25 frames/s, for at least six minutes at each location in the channel, and transferred simultaneously to a computer.  Visual inspection of the video movies clearly revealed long-term contact between particles, with rubbing and rolling interactions predominating.  Each video frame was analyzed, using standard image analysis techniques, to determine the centroids of the bright spots (Fig.~1b).  The velocity components $c_x$ and $c_y$ of the particles were determined from the displacement of the centroids between adjacent frames.  To ensure that spots in adjacent frames were from the same particle, the velocity of a particle was determined only if its displacement was less than $d_p/2$, limiting the maximum measurable velocity to roughly 10$\,$mm/s.  All measurements were made far enough from the entry and exit so that the flow was fully developed, i.e. the velocity field did not vary in the direction of flow.
	
\begin{figure}
\parbox[t]{0.25\textwidth}{\raggedright{\bf (a)}\\
\vspace*{-1.5em}
\centerline{\includegraphics[width=0.25\textwidth]{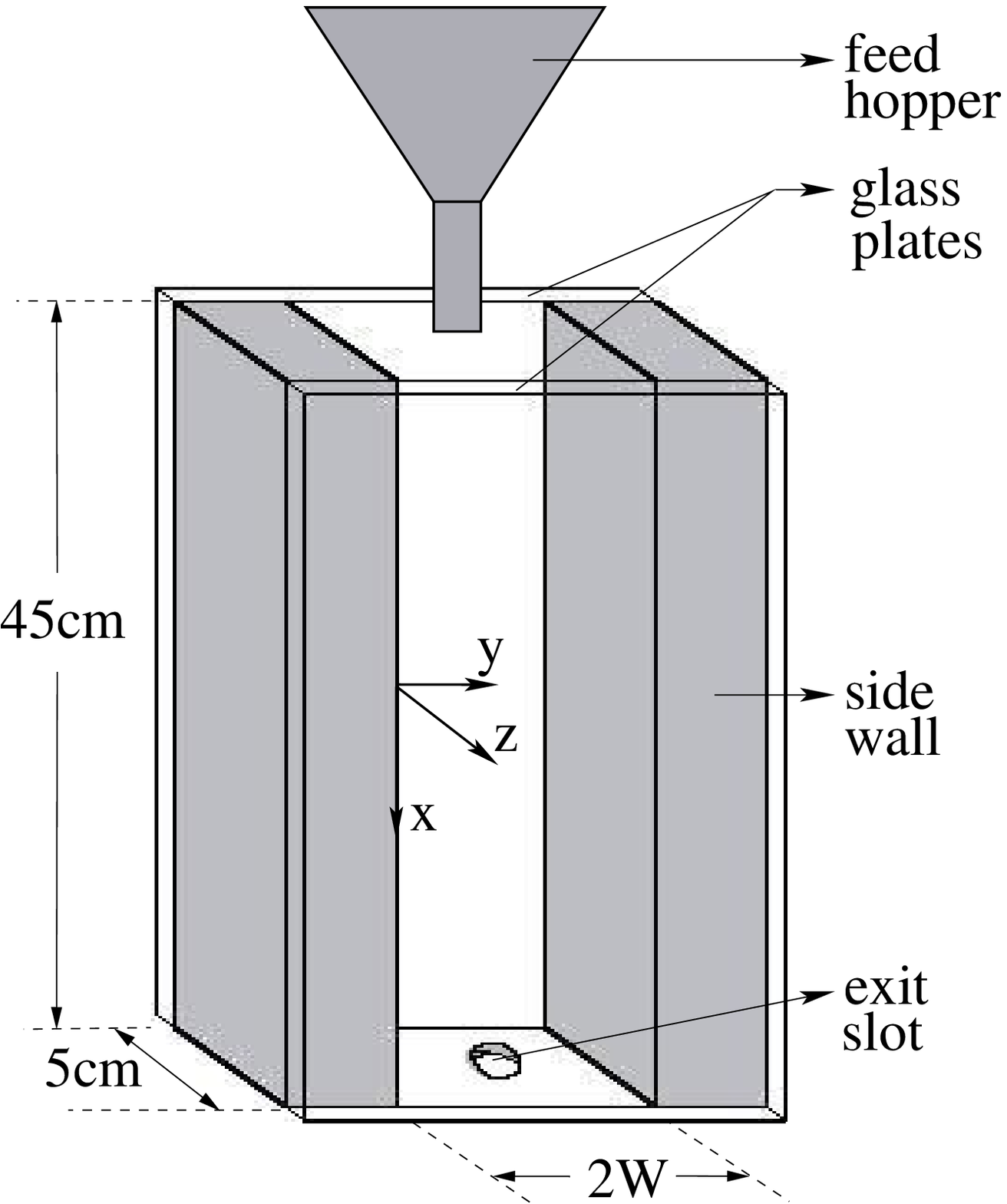}}} \hspace*{3ex}
\parbox[t]{0.15\textwidth}{\vspace*{6em}
\hspace*{-16ex} {\bf (b)}\\
\centerline{\includegraphics[width=0.15\textwidth]{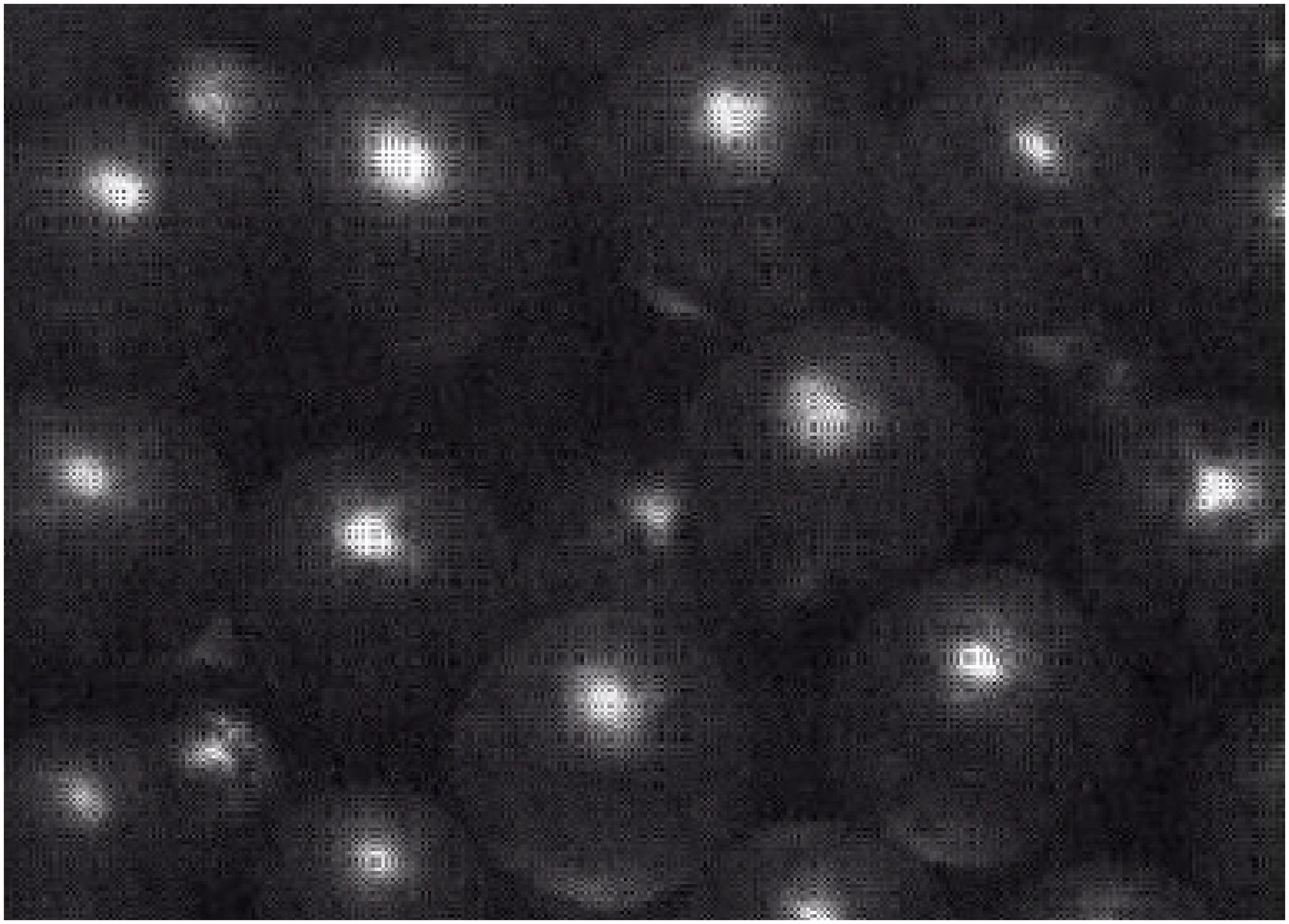}}}\\
\vspace*{0.5em}
\psfrag{v}{\boldmath $u_x$}
\psfrag{vv}{\boldmath $v$}
\psfrag{yy}[bc]{\boldmath \rule{4ex}{0pt}$u_x, \; v$ (mm/s)}
\psfrag{xx}[ct]{\boldmath $y/W$}
\centerline{\includegraphics[width=0.3\textwidth]{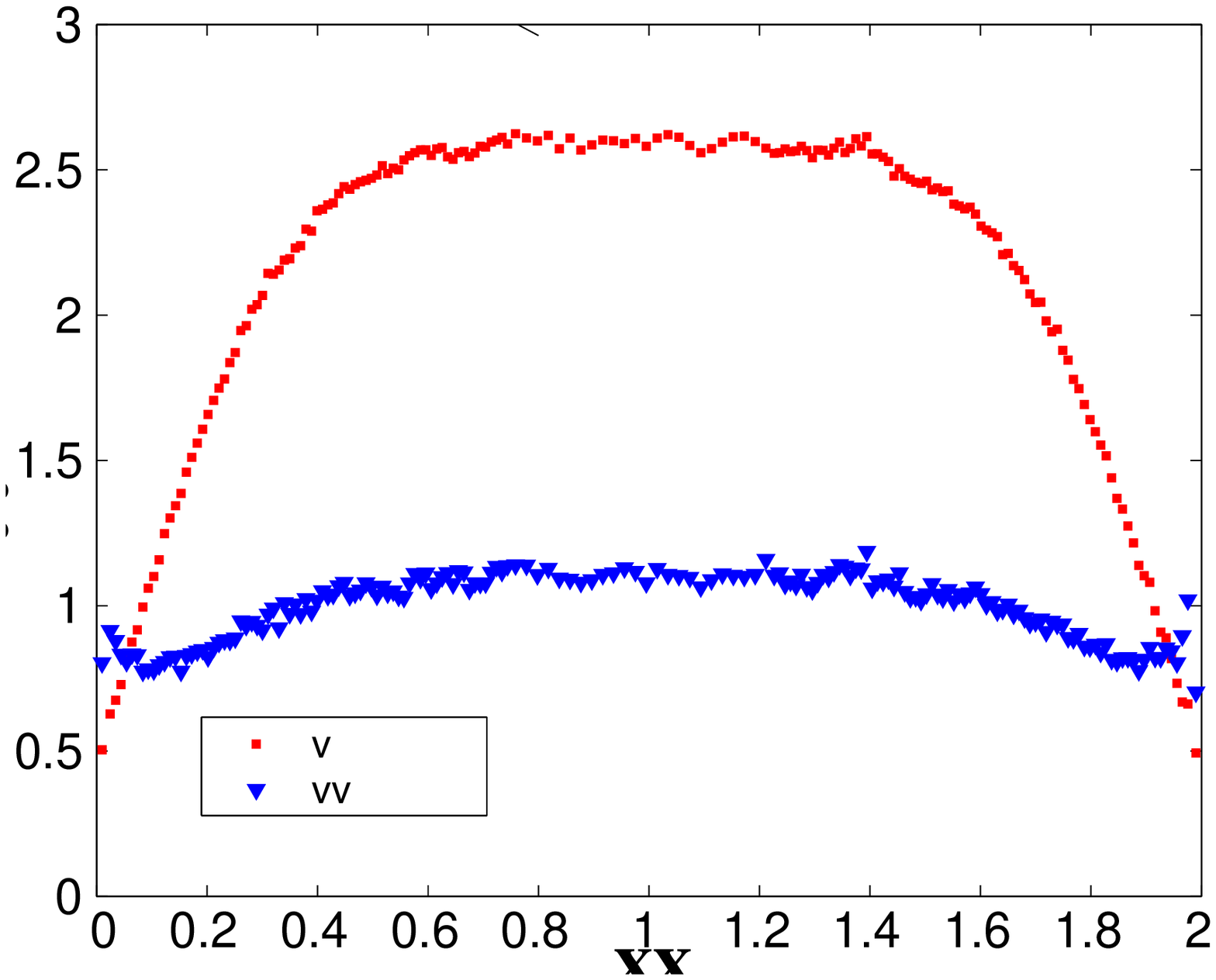}}
\vspace*{-12.5em}
\hspace*{-27ex} {\bf (c)}\\
\vspace*{12em}
\caption{
(color online) (a) A sketch (not to scale) showing the construction and dimensions of the channel.
(b) A snapshot of the flowing glass beads, showing the bright spots of reflected light.
(c) Profiles of the mean velocity $u_x$, and the root-mean-square velocity fluctuation $v$ for a rough walled channel with $W/d_p \! = \! 31.7$.\label{fig1}}
\end{figure}

	The profiles of mean properties were determined by notionally dividing the channel into vertical bins of width $d_p$ and averaging the relevant particle properties in each bin over the entire duration of the video movie.  Figure~1c shows a typical result for the mean velocity $u_x$ and the root-mean-square (rms) velocity fluctuation $v$.  The mean velocity increases sharply within the shear layer, from a small but finite slip at the wall to the maximum value $u_0$, and remains constant at $u_0$ in the substantial core.  This feature of shear banding is in agreement with previous work on flow in channels \cite{nedderman_laohakul80,natarajan_etal95} and cylindrical Couette cells \cite{mueth_etal00,losert_etal00}.  The dependence of the shear layer thickness on $W$ and the material and wall properties, and a comparison of the observed mean velocity field with predictions of available hydrodynamic theories will be reported elsewhere \cite{moka_etal04}.  Though the shear rate is negligibly small in the core, we find that $v$ is highest there (Fig.~1c); this is a surprising observation, as we expect the conversion of energy from mean to fluctuational motion to be high in regions where the shear rate is high.  
When a liquid (or a granular medium in the rapid flow regime \cite{haff83}) is sheared, 
the shear rate is high in regions where the temperature is high, as its viscosity decreases with rise in the temperature.  Our observation of the opposite here illustrates the difference in the micromechanics between dense frictional flows and molecular or rapid granular flows: the velocity fluctuations here arise from complex, collective motion, with particles contacts being sustained almost continuously, and not from particles flying between collisions with high velocity.

\begin{figure}
\begin{center}
\psfrag{X}[cb]{\boldmath $\xi_x$}
\psfrag{Y}[cb]{\boldmath $f_x(\xi_x)$}
\psfrag{  35.00}{\scriptsize \hspace*{-2.5ex} \begin{picture}(6,6) \color{red} \thicklines \put(2.5,2.5){\circle{2.5}} \end{picture} \hspace*{0.5ex} \raisebox{0.1ex}{23.4}}
\psfrag{  50.00}{\scriptsize \hspace*{-2.5ex} \begin{picture}(6,6) \color{blue} \thicklines \put(2.5,2.5){\circle{2.5}} \end{picture}  \hspace*{0.5ex} \raisebox{0.05ex}{31.7}}
\psfrag{  75.00}{\scriptsize \hspace*{-2.5ex} \begin{picture}(6,6) \color{dgreen} \thicklines \put(2.5,2.5){\circle{2.5}} \end{picture}  \hspace*{0.5ex} \raisebox{-0.05ex}{46.8}}
\psfrag{  100.00}{\scriptsize \hspace*{-2.5ex} \begin{picture}(6,6) \color{gold} \thicklines \put(2.5,2.5){\circle{2.5}} \end{picture}  \hspace*{0.5ex} \raisebox{-0.05ex}{64}}
\includegraphics[width=0.425\textwidth]{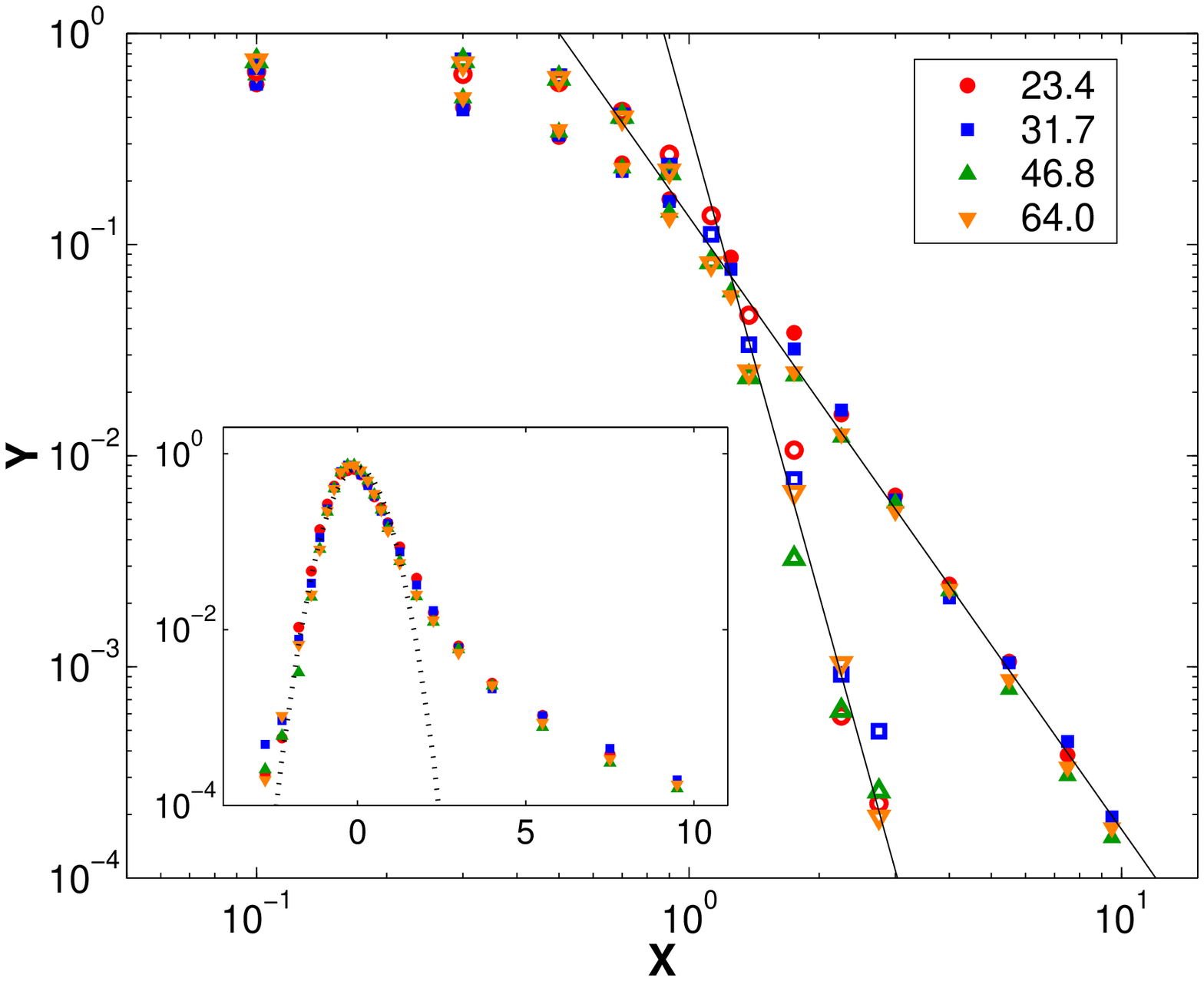}\\
\vspace*{-16em}
\hspace*{-36ex} {\bf (a)}\\
\vspace*{15.5em}
\psfrag{X}[cm]{\boldmath $\xi_y$}
\psfrag{Y}[cb]{\boldmath $f_y(\xi_y)$}
\hspace*{1ex} \includegraphics[width=0.425\textwidth]{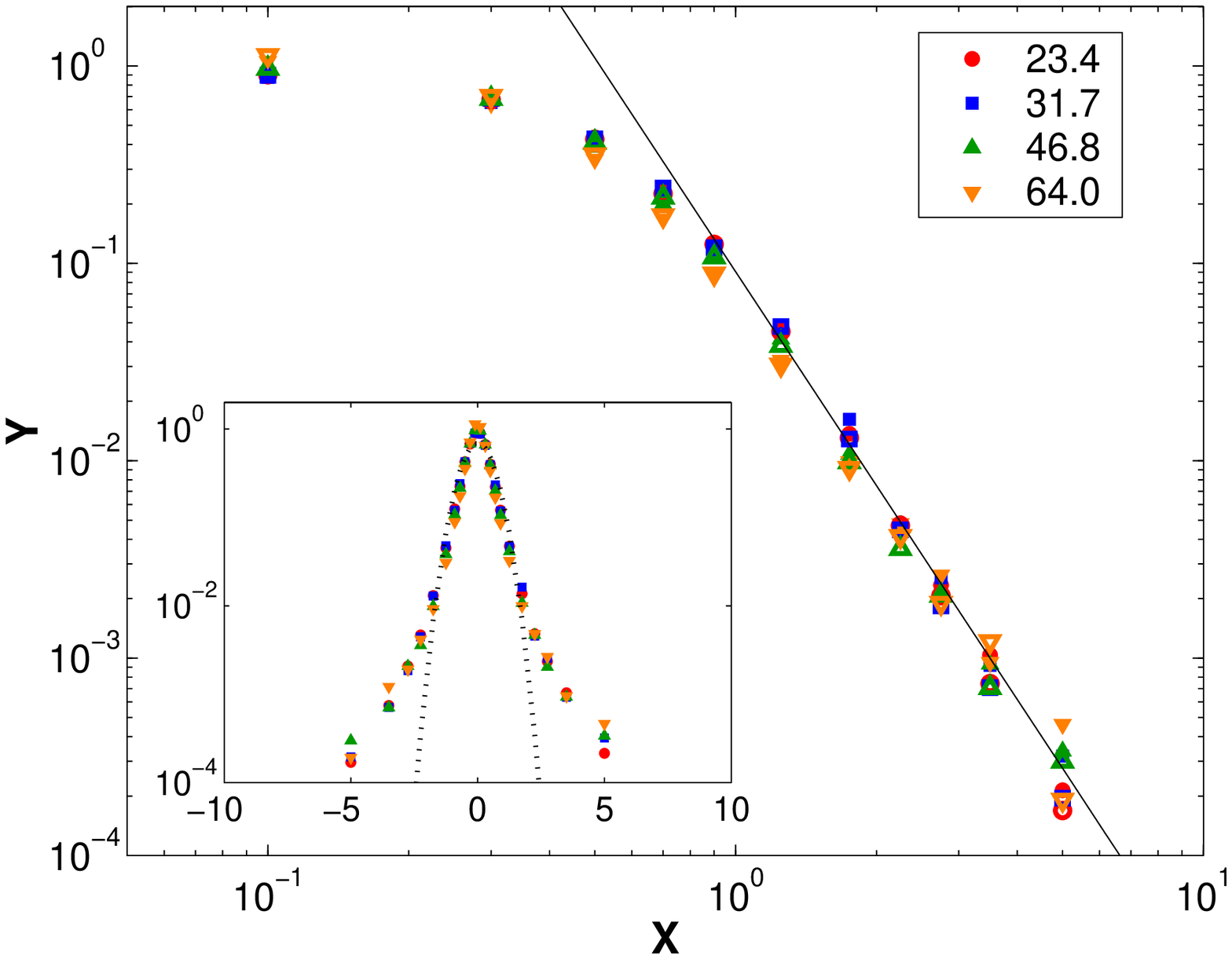}\\
\vspace*{-16em}
\hspace*{-36ex} {\bf (b)}\\
\vspace*{15em}
\caption{(color online) The probability distribution function for vertical (a) and horizontal (b) particle velocity in the fluid-like shear layer.  The data are for four different channel widths with rough walls (see text), the legend giving the ratio $W/d_p$.  The unfilled symbols in the main panels represent data for negative velocity fluctuations, reflected onto the positive side for accommodation in the log-log plot; the filled symbols represent data for positive velocity fluctuations.  The inset in each panel shows the distribution in linear-log axes.  The dotted line in the inset is the Gaussian distribution with the same variance.  The straight lines in the main panels are power-law fits for large velocity magnitude.\label{fig2}}
\end{center}
\end{figure}

	The distribution of velocities in the $x$ and $y$ directions (vertical and horizontal directions, respectively, see Fig.~1a) were determined as a function of position in the channel for a range of the channel width, with rough and smooth side walls.  Our visualization technique could not resolve movement in the $z$ direction.  The probability distribution function $f_x(\xi_x)$, where $\xi_x \equiv (c_x - u_x)/v$ is the dimensionless velocity fluctuation in the vertical direction, was determined by making a histogram of the number distribution of $\xi_x$, and normalizing it so that $\int \!\! f_x(\xi_x) d\xi_x = 1$; the distribution function $f_y(\xi_y)$ for horizontal velocities was determined similarly.  Figure~2 shows the distribution function in the shear layer for rough walled channels of four different channel widths.  We note that $f_x$ is not symmetric about $\xi_x=0$ (Fig.~2a), as gravity induces a preference for downward velocities.  Its decay is more rapid for negative $\xi_x$ than for positive.  There is no preferred direction for horizontal velocities and the distribution is therefore symmetric about $\xi_y=0$ (Fig.~2b).  In both directions, the distribution function decays as a power law, $f_i(\xi_i) \sim |\xi_i|^{-n}$, when $|\xi_i|$ is sufficiently large.

	A striking feature of Fig.~2 is the collapse of the data for all the channel widths into a single curve.  We find that data at other locations in the shear layer and for smooth walls (not shown) also collapse onto the same curve.  As the flow rate, and therefore the shear rate, varied between all these experiments, this implies that $f_x$ and $f_y$ do not depend on the shear rate, but only on the local value of $v$.  This result is further reinforced when we compare the data in the shear layer and the core (Fig.~3); the distribution function is identical in the two regions.  This is a remarkable result, as the nature of the flow in the two regions is very different - there is continuous deformation in the shear layer, but none in the core.  Curiously, this is similar to molecular fluids at thermodynamic equilibrium, where the Gaussian distribution holds regardless of the state of aggregation.  However, the distribution function away from equilibrium in molecular fluids depends on gradients of the velocity and temperature \cite{chapman_cowling}. Granular flows are fundamentally far from equilibrium (indeed the flowing state bears no resemblance to the equilibrium, static, state), and hence the shear-rate independence is a significant deviation from the statistical behavior of classical fluids.

	Our results are consistent with those of Ref.~\cite{choi_etal04}, who observed anisotropy and non-Gaussian distribution of the particle displacement over a time scale of 1 ms.  However, they found the distribution function changing to a Gaussian on coarse graining in time, but we find it distinctly non-Gaussian even with though our camera speed was 40 times slower.  We found no perceptible change in the distribution function when a faster video camera (60 frames/s) was used \cite{moka_etal04}, suggesting that our data is robust and valid over a range of time scales.
	
\begin{figure}
\begin{center}
\psfrag{X}[cm]{\boldmath $\xi_x$}
\psfrag{Y}[cb]{\boldmath $f_x(\xi_x)$}
\psfrag{  50.00s}{\scriptsize \hspace*{-2.5ex} \begin{picture}(6,6) \color{red} \thicklines \put(2.5,2.5){\circle{2.5}} \end{picture} \hspace*{-0.25ex} \raisebox{0.1ex}{31.7s}}
\psfrag{  50.00c}{\scriptsize \hspace*{-2.5ex} \begin{picture}(6,6) \color{blue} \thicklines \put(2.5,2.5){\circle{2.5}} \end{picture}  \hspace*{-0.25ex} \raisebox{0.05ex}{31.7c}}
\psfrag{  100.00s}{\scriptsize \hspace*{-2.5ex} \begin{picture}(6,6) \color{dgreen} \thicklines \put(2.5,2.5){\circle{2.5}} \end{picture}  \hspace*{-0.25ex} \raisebox{-0.05ex}{64s}}
\psfrag{  100.00c}{\scriptsize \hspace*{-2.5ex} \begin{picture}(6,6) \color{gold} \thicklines \put(2.5,2.5){\circle{2.5}} \end{picture}  \hspace*{-0.25ex} \raisebox{-0.05ex}{64c}}
\includegraphics[width=0.42\textwidth]{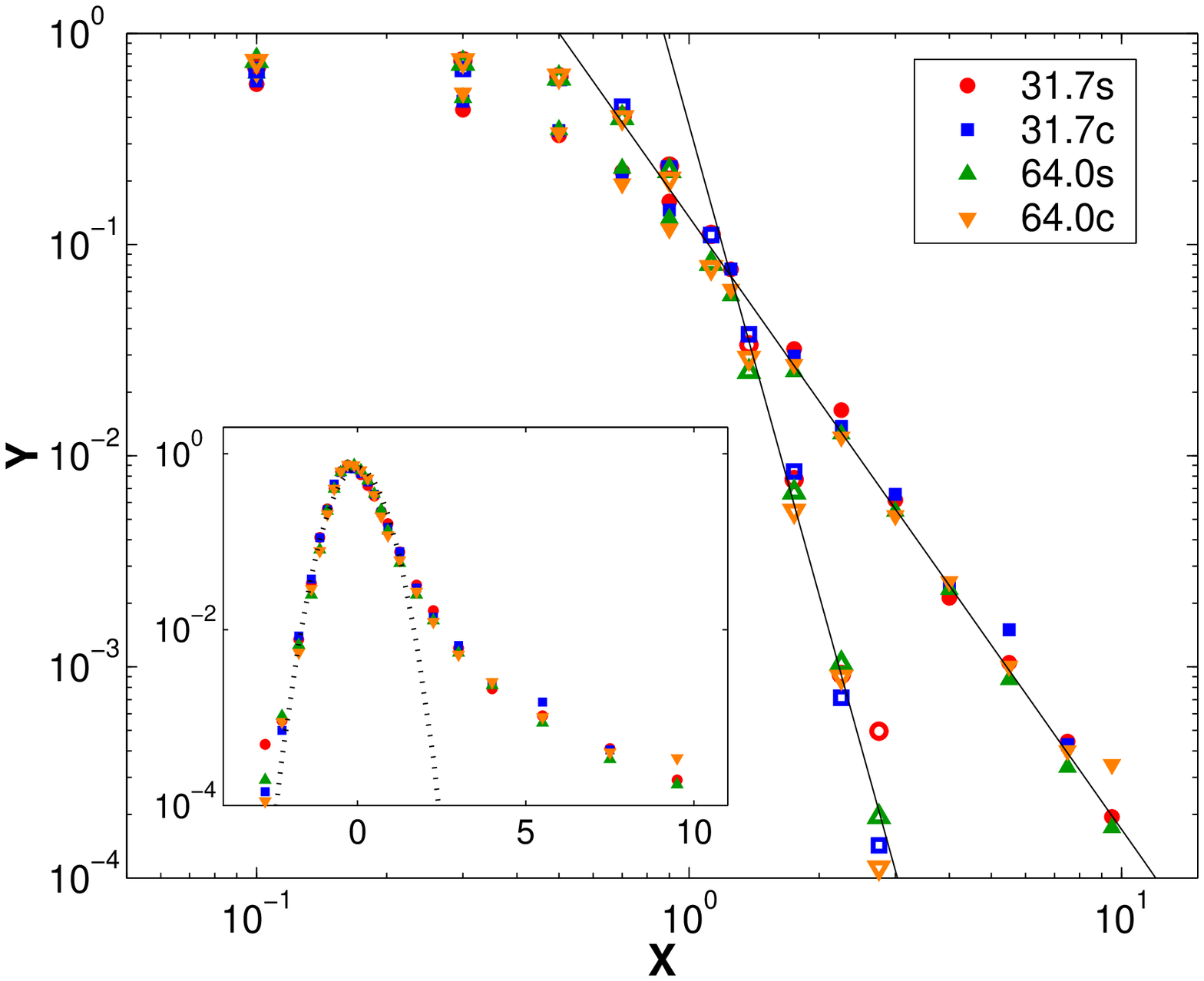}\\
\vspace*{-16em}
\hspace*{-36ex} {\bf (a)}\\
\vspace*{15.5em}
\psfrag{X}[cm]{\boldmath $\xi_y$}
\psfrag{Y}[cb]{\boldmath $f_y(\xi_y)$}
\includegraphics[width=0.42\textwidth]{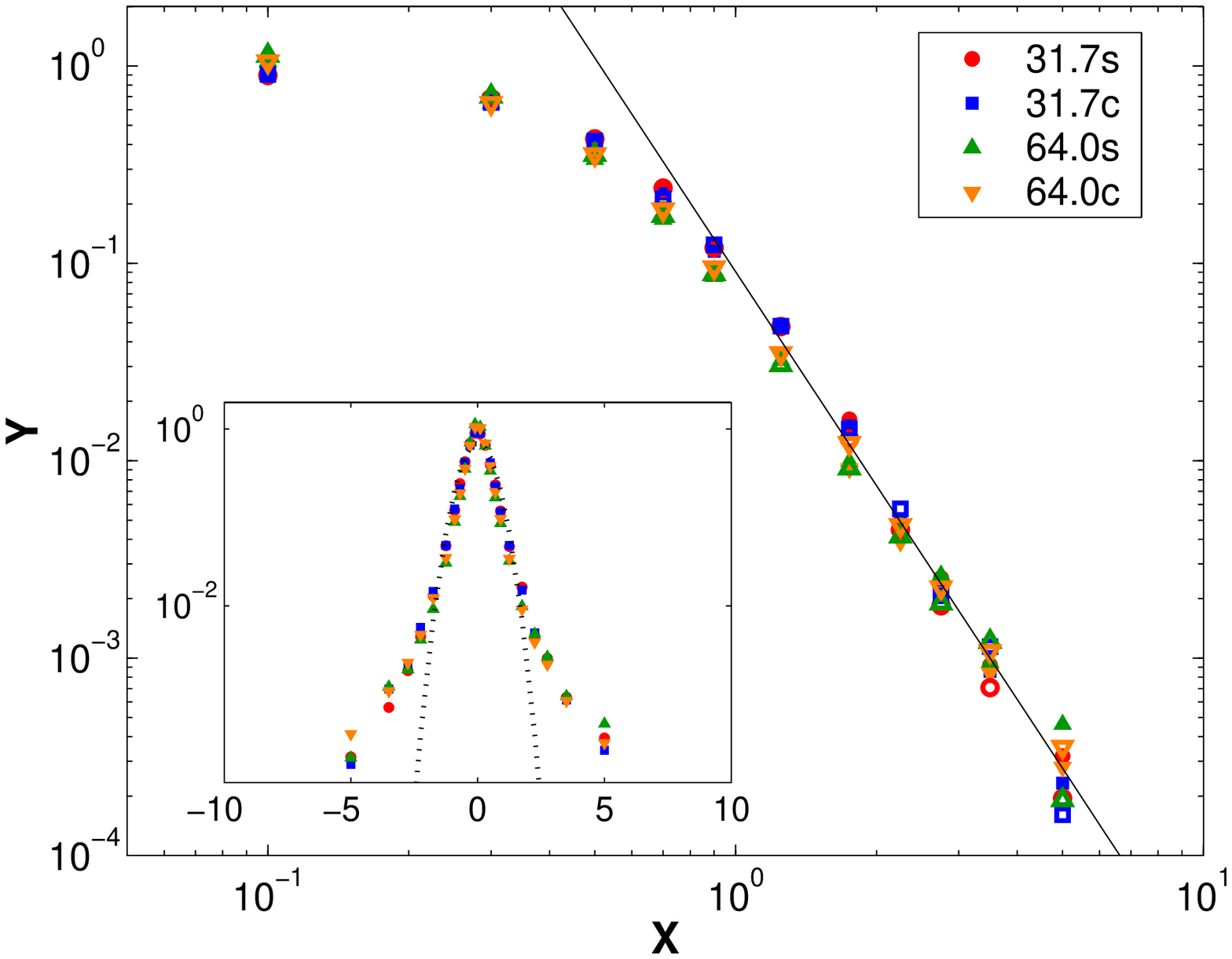}\\
\vspace*{-16em}
\hspace*{-36ex} {\bf (b)}\\
\vspace*{14em}
\vspace{1em} \caption{(color online) Comparison of the probability distribution function in the shear layer with that in the core.  The layout and symbols are as in Fig.~2.  The suffix `s' and `c' in the legend indicate data for the shear layer and the core, respectively.\label{fig3}}
\end{center}
\end{figure}

	The power-law exponents were determined by fitting a straight line through the data in the log-log plots for large velocities (Figs~2 and 3).  For the horizontal velocity the exponent $n_y$ is 3.6, and for the vertical velocity the exponents are $n_x^{+} \! = \! 2.9$ for positive $\xi_x$ and $n_x^{-} \! = \! 7.4$ for negative $\xi_x$.  The faster decay of the distribution for negative $\xi_x$ is a reflection of the preference for positive (downward) velocity fluctuations that gravity induces.  We do not, however, expect this form of the distribution to persist to very large velocities for the following reason: if we consider the mean velocity fluctuation square, or the specific fluctuational kinetic energy \cite{notes1}
$
v_{\infty}^2 = \int_{-\infty}^{\infty} \!\! (c_x - u_x)^2 \, f(c_x) dc_x + \int_{-\infty}^{\infty} \!\! c_y^2 \, f(c_y) dc_y,
$
we find that the first integral is unbounded because $n_x^{+} < 3$.  Thus, for the mean kinetic energy of particles to be finite, there has to be a faster decay of $f_x$ for large positive $c_x$.  This can be verified if higher particle velocities are measured, perhaps by high speed video imaging \cite{rouyer_menon00,choi_etal04}.

	The presence of velocity fluctuations even in the absence of a shear rate has been established in earlier studies \cite{menon_durian97,choi_etal04} on dense granular flows.  Our observation that $v$ is highest in the core is somewhat more surprising, as hydrodynamic theories \cite{johnson_etal90} give the transfer of kinetic energy from mean to fluctuational motion as the product of the shear stress and the shear rate, predicting that $v$ is minimum at the core.  A plausible explanation for the production of velocity fluctuations in the absence of shear is provided by considering the spatial correlation of the velocity fluctuation, defined as
$ C_y(Y) \equiv \langle \overline{c_x'}(t,y) \, \overline{c_x'}(t,y+Y) \rangle$.
Here $\overline{c_x'}(t,y)$ is the instantaneous fluctuation velocity averaged over particles in the bin centered at the reference point $y$, $\overline{c_x'}(t,y+Y)$ is the same in the bin centered at $y+Y$, and the angle brackets indicate an average over time.  Figure~4a shows that $C_y(Y)$ decays rapidly with $Y$, in the shear layer and the core, but not to zero.  There is a finite correlation at large distances, implying a correlated solid-like motion of the particles superimposed over the uncorrelated fluid-like motion.  We believe that this solid-like motion is a result of the stick-slip \cite{budny79,nasuno_etal98}, or jamming-unjamming, phenomena in granular materials.  When the medium is released, or unjammed, during the slip phase, the unbalanced force accelerates the particles for a short period of time, and a part of the kinetic energy acquired by the particles is retained in fluctuational modes.  The energy is lost in the next stick or jammed phase, but the cycle is repeated regularly.  The decay of the time correlation, $ C_t(T) \equiv \langle \overline{c_x'}(t,y) \, \overline{c_x'}(t+T,y) \rangle$,
to zero in Fig.~4b indicates that the stick-slip motion itself is uncorrelated beyond a time scale of roughly half a second.  (Note that $C_t(T)$ is not the usual autocorrelation function, as we are not tracking the velocity of individual particles in time).  The relative unimportance of shear in generating fluctuational motion implies that shear work largely dissipates as true heat through rubbing and rolling friction \cite{johnson_etal90}.   It appears likely that the jamming-unjamming cycle is affected by shear, but it is not clear why shear diminishes $v$ but leaves the velocity distribution unchanged.


\begin{figure}
\psfrag{x}[cm]{\boldmath $Y/d_p$}
\psfrag{y}[cb]{\boldmath \rule[-0.5em]{0ex}{0.5em} $C_y(Y)/C_y(0)$}
\centerline{\includegraphics[width=0.30\textwidth]{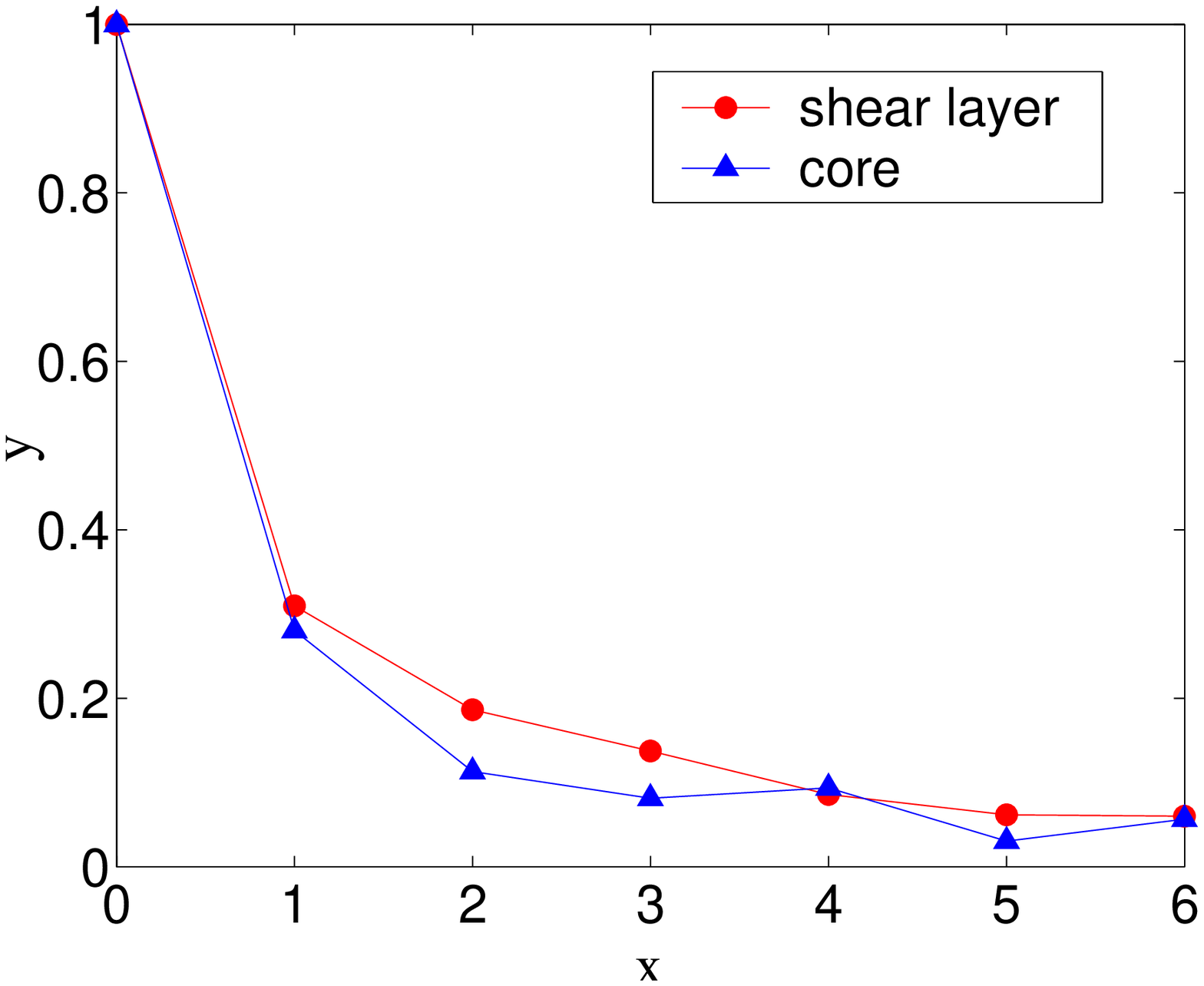}}
\vspace*{-18em}
\hspace*{-36ex} {\bf (a)}\\
\vspace*{17.5em}
\psfrag{x}[cm]{\boldmath $T (s)$}
\psfrag{y}[cb]{\boldmath $C_t(T)/C_t(0)$}
\centerline{\includegraphics[width=0.30\textwidth]{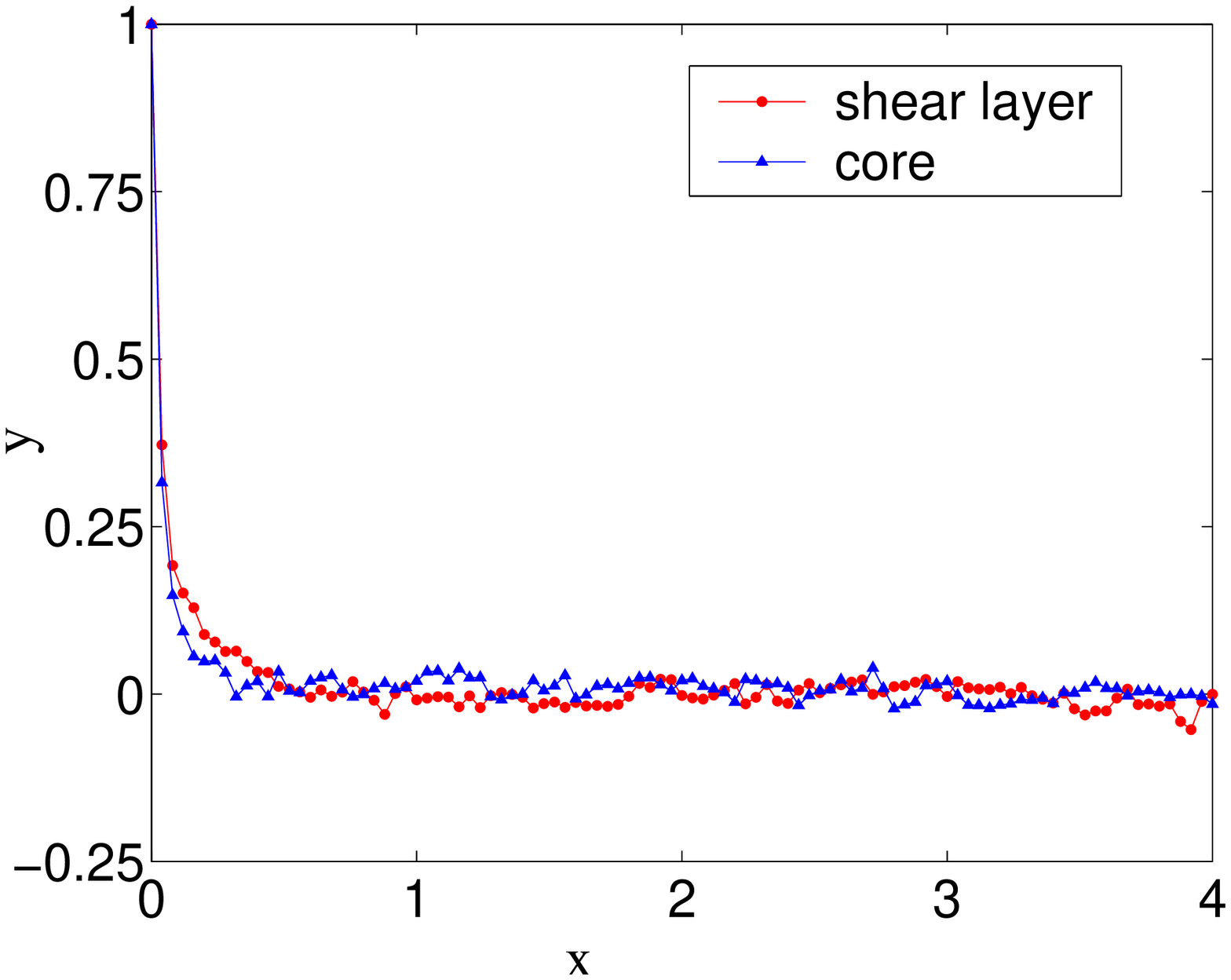}}
\vspace*{-17.5em}
\hspace*{-36ex} {\bf (b)}\\
\vspace*{16em}
\caption{(color online) Correlation functions of the velocity fluctuation in the shear layer and the core (see text for definitions of $C_y$ and $C_t$) for a rough walled channel with $W/d_p \! = \! 31.7$.  The reference point $y$ was $4 \, d_p$ from the left wall for the shear layer, and channel center for the core.  (a) The spatial correlation as a function of the horizontal separation $Y$, and (b) the time correlation as a function of the time delay $T$.\label{fig4}}
\end{figure}

	The above discussion leads to the important question of whether the velocity distribution we have reported is universal for slow granular flows, or is limited to flows driven by a body force. This is an issue of considerable importance, and can only be answered by conducting experiments in other geometries.  A kinematic variable that we have not measured in this study is the rotation velocity of the particles.  Recent studies that have modelled granular media as Cosserat continua \cite{muhlhaus_vardoulakis87,mohan_etal02} have shown the rotation velocity to play an important role in determining the kinematics.  Though its measurement is more difficult, it will be useful in critically evaluating hydrodynamic theories and in understanding the microscopic difference between the solid-like and fluid-like regions.

We thank K. S. Ananda for assistance in preparing the plots, and S. Ramaswamy and N. Menon for useful discussions.  This work was supported by the Department of Science and Technology, India.


\begin{thebibliography}{21}
\bibitem[{not()}]{notes1}
\bibinfo{note}{In writing this form for $v_{\infty}$, we have assumed that the
  $x$ and $y$ velocities are uncorrelated. This has been verified for
  velocities around the mean, but the data for large velocity fluctuations is
  too sparse to allow verification.}
\end{thebibliography}

\end{document}